\documentclass[pra,twocolumn,superscriptaddress]{revtex4-1}
\usepackage{amssymb}
\usepackage{graphicx}
\usepackage{dcolumn}
\usepackage{bm}
\usepackage{amsmath}
\usepackage[usenames]{color}
\usepackage{textcomp}
\usepackage{subfigure}
\usepackage{booktabs}
\usepackage{diagbox}

\usepackage[colorlinks,linkcolor=blue,anchorcolor=blue,citecolor=blue,urlcolor=blue,driverfallback=dvipdfm]{hyperref}

\begin{document}

\title{Nonequilibrium magnetic dynamics of the two-component Bose-Hubbard model}
\author{Hui Tan}
\affiliation{Department of Physics, National University of Defense Technology, Changsha 410073, P. R. China}
\author{Jianmin Yuan}
\affiliation{Institute of Atomic and Molecular Physics, Jilin University, Changchun 130012, P. R. China}
\affiliation{Department of Physics, National University of Defense Technology, Changsha 410073, P. R. China}
\author{Yongqiang Li}
\email{li\_yq@nudt.edu.cn}
\affiliation{Department of Physics, National University of Defense Technology, Changsha 410073, P. R. China}
\affiliation{Hunan Key Laboratory of Extreme Matter and Applications, National University of Defense Technology, Changsha 410073, P. R. China}
 \affiliation{Hunan Research Center of the Basic Discipline for Physical States, National University of Defense Technology, Changsha 410073, China}

\begin{abstract}
A central challenge in strongly interacting many-body systems is understanding the far-from-equilibrium dynamics. Here, we study the many-body magnetic dynamics of the two-component Bose-Hubbard model by
developing a two-component extension of nonequilibrium bosonic dynamical mean-field theory.
Using this numerical method, we uncover rich quantum spin dynamics via inter-species interaction quenches. A sudden ramp-up of interactions induces slow thermalization, leading to a long-lived metastable state, whereas quenching to weak interactions results in rapid thermal equilibrium, featuring a two-step relaxation behavior through distinct exponential decays. Furthermore, under periodic modulation of the inter-species interactions, emergent Floquet dynamics drives a transition from a magnetic to an unordered phase.
\end{abstract}

\date{\today}
\maketitle

\affiliation{Department of Physics, National University of Defense
Technology, Changsha 410073, P. R. China}

\affiliation{Department of Physics, National University of Defense
Technology, Changsha 410073, P. R. China}

\affiliation{Hefei National Laboratory for Physical Sciences at the Microscale and Department of Modern Physics,
     University of Science and Technology of China, Hefei 230026, China}
\affiliation{CAS Center for Excellence in Quantum Information and Quantum Physics,
     University of Science and Technology of China, Hefei 230026, China}

\affiliation{Department of Physics, Graduate School of China Academy of
Engineering Physics, Beijing 100193, P. R. China}

\affiliation{Department of Physics, National University of Defense
Technology, Changsha 410073, P. R. China}


\section{\label{sec:level1}introduction}
Ultracold bosonic atoms in optical lattices offer unprecedented dynamical tunability, establishing them as ideal platforms for investigating nonequilibrium dynamics~\cite{trotzky_probing_2012,preiss2015strongly,eisert_quantum_2015,langen2015ultracold,RevModPhys.80.885,gross2017quantum}. Extensive experimental studies over the past few years have revealed exotic nonequilibrium phenomena in the bosonic systems, such as quench dynamics~\cite{greiner2002collapse,10.1126/science.1192368}, spin dynamics~\cite{PhysRevLett.95.190405,trotzky2008time}, many-body localization~\cite{10.1126/science.aaf8834}, Floquet dynamics in driven systems~\cite{PhysRevLett.116.205301,PhysRevX.10.021044}, Kibble-Zurek mechanism near quantum phase transitions~\cite{PhysRevLett.106.235304,10.1073/pnas.1408861112,PhysRevLett.127.200601}, and Hilbert-space fragmentation in constrained systems~\cite{PhysRevLett.133.196301,adler_observation_2024,10.1126/sciadv.adv3255}. These experimental achievements have attracted significant theoretical attention~\cite{PhysRevA.90.033606,PhysRevLett.98.180601,PhysRevLett.106.095702,PhysRevA.84.033620,dziarmaga_quench_2014,PhysRevB.86.085140,RevModPhys.89.011004,PhysRevX.5.011038,Sierant_2025,PhysRevLett.105.220401,Moudgalya_2022,RevModPhys.83.863,RevModPhys.78.179,kennett_out--equilibrium_2013,JAKSCH200552,RevModPhys.93.025003,ueda2020quantum,RevModPhys.91.021001}. Nevertheless, these prior theoretical researches have largely concentrated on the dynamics of charge degrees of freedom or on lower spatial dimensions, modeled effectively by the single-species Bose-Hubbard model. 

Two-component bosonic mixtures in optical lattices serve as a versatile testbed for studying quantum spin dynamics, particularly in higher dimensions. For the system, its ground-state properties have been extensively studied, revealing that inter-species interactions give rise to a plethora of rich quantum phases, including counterflow superfluidity~\cite{altman_phase_2003,PhysRevLett.90.100401,zheng_counterflow_2025}, non-integer Mott insulator~\cite{PhysRevLett.125.245301}, and various magnetic ordering~\cite{altman_phase_2003,PhysRevB.80.245109,PhysRevB.84.144411,PhysRevA.85.023624,li_anisotropic_2013}. However, simulating the real-time nonequilibrium dynamics of the two-component Bose-Hubbard model is computationally challenging in higher dimensions~\cite{PhysRevA.109.013308,PhysRevA.111.023324}, as a result of the exponential growth of Hilbert space dimensionality. Various large-scale numerical methods are implemented to tackle these complex dynamical problems. For example, density-matrix renormalization group methods have demonstrated accurate predictions for dynamical results in one-dimensional optical lattices~\cite{RevModPhys.77.259}, yet they face limitations in higher dimensions due to rapid entanglement growth. The time-dependent variational Monte Carlo approach is well-suited for investigating correlation spreading in finite systems but complicates the study of thermalization ~\cite{PhysRevA.89.031602,PhysRevLett.105.250401}. In higher dimensions, although Gutzwiller mean-field theory offers a simple means of describing nonequilibrium phenomena, it falls short in capturing magnetic phases with spin ordering~\cite{PhysRevB.75.085106,PhysRevLett.105.220401,PhysRevA.109.013308}.

An alternative method to describe the dynamics of correlated many-body systems in higher dimensions is  
nonequilibrium dynamical mean-field theory. Analogy to equilibrium, the nonequilibrium version of dynamical mean-field theory transforms a lattice model with local interactions into an effective impurity model that is suitable for numerical solutions. The impurity problem can be solved within the Keldysh formalism~\cite{Stefanucci_van_Leeuwen_2013,RevModPhys.86.779,PhysRevA.111.023324}, which is based on nonequilibrium Green’s functions, and provides a robust approach to characterize the rapid transient evolution and nonequilibrium steady states of many-body systems by illustrating the movement of particles and holes across various space-time positions. When reformulated using the Keldysh formalism, the impurity effective action is defined on the three-branch Kadanoff-Baym contour, and can be addressed through weak-coupling expansions~\cite{PhysRevB.88.165115,PhysRevLett.110.136404}, strong-coupling expansions~\cite{PhysRevLett.105.146404,PhysRevB.82.115115}, quantum Monte Carlo~\cite{PhysRevLett.105.096402,eckstein_thermalization_2009,tsuji_dynamical_2011}, and Hamiltonian-based impurity solvers~\cite{PhysRevB.88.235106,PhysRevB.90.235131,PhysRevB.91.045136}.
This versatility enables dynamical mean-field theory to elucidate  nonequilibrium phenomena in both homogeneous~\cite{PhysRevLett.103.056403,PhysRevLett.97.266408,PhysRevLett.107.186406,PhysRevLett.105.146404,PhysRevB.86.205101,PhysRevLett.110.136404} and inhomogeneous systems~\cite{PhysRevB.88.075135,PhysRevLett.123.193602,PhysRevB.108.125143}. Presently, nonequilibrium dynamical mean-field theory has already been extended to the bosonic cases (BDMFT), and to examine quench dynamics of the single-component Bose-Hubbard model using a strong-coupling perturbative impurity solver~\cite{PhysRevX.5.011038}, $i.e.$ lowest-order noncrossing approximation (NCA). Nevertheless, a multi-component extension of nonequilibrium BDMFT remains unexplored. 

In this work, we implement a two-component extension of the nonequilibrium single-component BDMFT formalism,
based on the NCA impurity solver~\cite{PhysRevX.5.011038}. We first benchmark our approach against the single-component Bose-Hubbard model, demonstrating excellent agreement with existing results for both ground-state properties and far-from-equilibrium dynamics.
For the two-component Bose-Hubbard model, the ground-state magnetic phase diagram matches well with the results obtained from the Hamiltonian-based exact diagonalization (ED) solver. Then, we explore the quench dynamics of the two-component Bose-Hubbard model. Specifically, when the inter-species interactions are rapidly increased, the system undergoes slow thermalization, stabilizing a long-lived metastable state. On the contrary, when the inter-species interactions are quenched to zero, 
the system rapidly thermalizes via a two-step exponential relaxation to equilibrium. Additionally, under periodic modulation of inter-species interactions, we observe a magnetic-to-unordered crossover in the Floquet-driven steady states.

The paper is structured as follows: Section~\ref {sec:level2} presents a detailed description of the two-component Bose-Hubbard model and our approach. Section~\ref{sec:level3} discusses the results of our model on the equilibrium magnetic phase diagram and nonequilibrium dynamics. In section~\ref{sec:level4}, we provide a summary and discussion.

\section{\label{sec:level2} Model and Method} We focus on the two-component Bose-Hubbard model, which can be implemented by utilizing a Bose-Bose mixture of two different species~\cite{PhysRevA.77.011603,PhysRevLett.100.210402} or two different internal states of a single species~\cite{PhysRevLett.105.045303,PhysRevX.9.041014} loaded into optical lattices. For a sufficiently deep lattice and low filling, the corresponding model is given by~\cite{PhysRevB.75.085106,PhysRevLett.105.220401},
\begin{eqnarray}{\label{Ham}}
	\hat{H}&=&-\sum_{\langle i,j \rangle}J_{\sigma}(\hat{b}^{\dagger}_{i,\sigma}\hat{b}_{j,\sigma} + \mathrm{h.c.}) + \sum_{i,\sigma\neq\sigma'}U_{\sigma \sigma'}\hat{n}_{i,\sigma}\hat{n}_{i,\sigma'}\nonumber\\
	&+&\sum_{i,\sigma}\frac{U_{\sigma}}{2}\hat{n}_{i,\sigma}(\hat{n}_{i,\sigma}-1) - \sum_{i,\sigma}\mu_{\sigma}\hat{n}_{i,\sigma},
\end{eqnarray}
where $\langle i,j \rangle$ stands for nearest-neighbour sites $i$ and $j$, $\hat{b}^{\dagger}_{i,\sigma}$ ($\hat{b}_{i,\sigma}$)
denotes the bosonic creation (annihilation) operator for species $\sigma$ at site $i$, and $\hat{n}_{i,\sigma}=\hat{b}^\dagger_{i,\sigma}\hat{b}_{i,\sigma}$ is the local
density. $J_{\sigma}$ is the hopping amplitudes for species $\sigma=1,2$. Generally, the two species tunnel with unequal amplitudes as a result of different
masses or a spin-dependent optical lattice. Without loss of generality, we set $J_1=J_2\equiv J$ as the unit of energy in our studies. $U_{\sigma\sigma'}$ are onsite interactions between species $\sigma$ and $\sigma^\prime$, which can be tuned via Feshbach resonances or spin-dependent lattices~\cite{PhysRevLett.92.160406,PhysRevLett.105.045303}. $\mu_{\sigma}$ is the chemical potential.

\subsection{Nonequilibrium bosonic dynamical mean-field theory}
To study the nonequilibrium many-body dynamics, we extend equilibrium bosonic dynamical mean-field theory using the Kadanoff-Baym formalism on an L-shaped contour. This extension enables the direct solution of the quantum impurity problem. By adapting the formalism in Ref.~\cite{PhysRevX.5.011038} through modifications to the local atom-atom interactions and the Nambu spinor, the effective action for the impurity site is obtained, 
\begin{eqnarray}
S_{\mathrm{eff}} &=& \int_{\mathcal{C}} \mathrm{d} t\left[-\sum_{\sigma}\mu_{\sigma} n_{\sigma}(t)+U_{\sigma\sigma'}(t)n_{\sigma}(t)n_{\sigma'}(t)\right]\nonumber\\
&+&\int_{\mathcal{C}} \mathrm{d} t\sum_{\sigma}\frac{U_{\sigma}}{2} n_{\sigma}(t)\bigg[n_{\sigma}(t)-1\bigg]\nonumber\\
&+& \int_{\mathcal{C}} \mathrm{d} t\left[-z J \boldsymbol{\Phi}^{\dagger}(t)-\int_{\mathcal{C}} \mathrm{d} t^{\prime} \boldsymbol{\Phi}^{\dagger}\left(t^{\prime}\right) \boldsymbol{\Delta}\left(t^{\prime}, t\right)\right] \mathbf{b}(t)\nonumber\\
&+& \frac{1}{2} \iint_{\mathcal{C}} \mathrm{d} t \mathrm{d} t^{\prime} \mathbf{b}^{\dagger}(t) \boldsymbol{\Delta}\left(t, t^{\prime}\right) \mathbf{b}\left(t^{\prime}\right),
\end{eqnarray}
where $z$ is the lattice coordination number. The Nambu notation is introduced with $\mathbf{b}^\dagger \equiv (b_{1}^{\dagger}\  b_{1}\  b_{2}^{\dagger}\  b_{2})$, where the site index for the bosonic creation (annihilation) operator $b_{\sigma}^{\dagger}$ ($b_{\sigma}$) is omitted for simplify. $\boldsymbol{\Delta}\left(t, t^{\prime}\right)$ is the two-time hybridization function, where the time arguments $t$ and $t'$ lie on the three branches of the Kadanoff-Baym contour $\mathcal{C}$. The contour integral $\int_{\mathcal{C}}$ is computed along $\mathcal{C}\equiv\mathcal{C}_1:0\rightarrow t_\mathrm{max}\cup\mathcal{C}_2:t_\mathrm{max}\rightarrow 0\cup\mathcal{C}_3:0 \rightarrow -i\beta$~\cite{RevModPhys.86.779}, where $t_\mathrm{max}$ is the maximum evolution time, and $\beta$ is the reciprocal of the temperature $T$. The superfluid order parameter is defined as $\boldsymbol{\Phi}^{\dagger}=(\langle b_{1}^{\dagger} \rangle\  \langle b_{1} \rangle\ \langle b_{2}^{\dagger} \rangle\  \langle b_{2} \rangle)$.
 
\subsection{Impurity solver based on the states propagator}
Usually in equilibrium BDMFT, the  effective action $S_{\mathrm{eff}}$ can be mapped to an Anderson impurity model and then solved by exact diagonalization or numerical renormalization group, namely Hamiltonian based approaches. Nevertheless, extending the Hamiltonian-based impurity solver to a nonequilibrium version is hard, due to the huge Hilbert space~\cite{PhysRevB.88.235106,PhysRevB.90.235131,PhysRevB.91.045136}.
Diagrammatic approaches provide an alternative route for handling the effective action $S_{\mathrm{eff}}$, employing techniques like continuous-time quantum Monte Carlo algorithms ~\cite{PhysRevLett.105.096402,eckstein_thermalization_2009,tsuji_dynamical_2011} and weak- or strong-coupling perturbation theory~\cite{PhysRevB.88.165115,PhysRevLett.110.136404,PhysRevLett.105.146404,PhysRevB.82.115115}. In this work, we solve the impurity action $S_{\mathrm{eff}}$ using strong-coupling perturbation theory, implementing a hybridization expansion on the contour $\mathcal{C}$. This strong-coupling expansion method has already been applied to both nonequilibrium fermionic and bosonic dynamical mean-field theory employing the pseudo-particle technique~\cite{PhysRevB.82.115115,PhysRevX.5.011038}. Here, we extend the method to the two-component bosonic case, and derive the strong-coupling formalism based on the states propagator~\cite{PhysRevB.87.075124,PhysRevLett.120.197601,RevModPhys.86.779}, which is defined by
\begin{eqnarray}
	\mathcal{P}(t,t')&=&-i\mathcal{T}_\mathcal{C}{\rm exp}\left[-i\int_{t'}^{t}\mathrm{d}t_a\mathcal{H}_{\rm loc}(t_a)\right]\\
 &\times&\mathrm{exp}\left[- \frac{i}{2} \int_{t'}^{t} \mathrm{d} t_a \mathrm{d} t_b \mathbf{b}^{\dagger}(t_a) \boldsymbol{\Delta}\left(t_a, t_b\right) \mathbf{b}(t_b)\right]\nonumber.
\end{eqnarray}
Here, $\mathcal{T}_\mathcal{C}$ is a contour-ordering operator, and
\begin{eqnarray}
\mathcal{H}_{\rm loc}(t_a)&=&-\sum_{\sigma}\mu_{\sigma} n_{\sigma}(t_a)+U_{\sigma\sigma'}n_{\sigma}(t_a)n_{\sigma'}(t_a)\\
&+&\sum_{\sigma}\frac{U_{\sigma}}{2} n_{\sigma}(t_a)\big[n_{\sigma}(t_a)-1\big]\nonumber\\ 
&-& \left[z J \boldsymbol{\Phi}^{\dagger}(t_a)+\int_{\mathcal{C}} \mathrm{d} t^{\prime} \boldsymbol{\Phi}^{\dagger}\left(t^{\prime}\right) \boldsymbol{\Delta}\left(t^{\prime}, t_a\right)\right] \mathbf{b}(t_a),\nonumber
\end{eqnarray}
where the integral $\int_{t'}^{t}$ is computed from $t'$ to $t$ along the contour $\mathcal{C}$.

Specifically, different types of $\mathcal{P}(t,t')$ are defined on the three branches of the contour $\mathcal{C}$, as shown in Table~\ref{tab:table1}, and hold the following symmetry relations,
\begin{eqnarray}
	\left(\mathcal{P}^{>(<)}\left(t, t^{\prime}\right) \right)^\dagger &=& -\mathcal{P}^{>(<)}\left(t^{\prime}, t\right),
\end{eqnarray}
\begin{eqnarray}
\left(\mathcal{P}^{\rceil}\left(t, \tau\right) \right)^\dagger &=& \mathcal{P}^{\lceil}\left(\beta-\tau, t \right).
\end{eqnarray}
\begin{table}[b]
\caption{\label{tab:table1}%
Various $\mathcal{P}(t,t')$ defined on the three branches of the contour $\mathcal{C}$. }
\begin{ruledtabular}
\renewcommand{\arraystretch}{1.3}
\begin{tabular}{ c|m{2.5cm}<{\centering} | m{2.5cm}<{\centering} | m{2.5cm}<{\centering} }
	\diagbox{$t$}{$t'$} & $\mathcal{C}_1$ & $\mathcal{C}_2$ & $\mathcal{C}_3$ \\ 
	\hline
	$\mathcal{C}_1$ & $\mathcal{P}^{++}\left(t, t^{\prime}\right)$ & $\mathcal{P}^{<}\left(t, t^{\prime}\right)$ & $\mathcal{P}^{\rceil}\left(t, \tau^{\prime}\right)$ \\ 
	\hline
	$\mathcal{C}_2$ & $\mathcal{P}^{>}\left(t, t^{\prime}\right)$ & $\mathcal{P}^{--}\left(t, t^{\prime}\right)$ & $\mathcal{P}^{\rceil}\left(t, \tau^{\prime}\right)$ \\   
	\hline
	$\mathcal{C}_3$ & $\mathcal{P}^{\lceil}\left(\tau, t^{\prime}\right)$ & $\mathcal{P}^{\lceil}\left(\tau, t^{\prime}\right)$ & $i \mathcal{P}^M\left(\tau-\tau^{\prime}\right)$ \\	
\end{tabular}
\end{ruledtabular}
\end{table}
Then, we can calculate the partition function
\begin{eqnarray}
	Z=i{\rm T_r}\left[\mathcal{P^{<}}(t,t) \right],
\end{eqnarray}
and the expectation value of operator $O(t)$
\begin{eqnarray}
	\langle O(t) \rangle=\frac{1}{Z}{\rm T_r}\left[i\mathcal{P^{<}}(t,t)O(t) \right].
\end{eqnarray}

We subsequently expand $\mathcal{P}\left(t, t^{\prime}\right)$ in powers of the hybridization-function part, and finally arrive at the Volterra integro-differential equation according to the diagrammatic rules~\cite{PhysRevB.87.075124,PhysRevLett.120.197601,RevModPhys.86.779},
\begin{eqnarray}\label{dyson}
	i \partial_t \mathcal{P}\left(t, t^{\prime}\right)&=&\mathcal{H}_{\mathrm{loc}}(t) \mathcal{P}\left(t, t^{\prime}\right)\nonumber\\
	&+&\int_{t^{\prime}}^t \mathrm{d} t_a \mathcal{S}\left(t, t_a\right) \mathcal{P}\left(t_a, t^{\prime}\right).
\end{eqnarray}
Expanding $\mathcal{P}\left(t, t^{\prime}\right)$ to the first order gives the noncrossing approximation for self-energy,
\begin{eqnarray}\label{Self}
	\mathcal{S}(t_b,t_a) &=&\frac{i}{2} \sum_{\alpha,\beta}\mathbf{b}_{\alpha}(t_b)\mathcal{P}(t_b,t_a)\mathbf{b}^{\dagger}_{\beta}(t_a)\boldsymbol{\Delta}_{\alpha\beta}\left(t_b, t_a\right)\nonumber\\
	&+& \frac{i}{2}\sum_{\alpha,\beta}\mathbf{b}_{\alpha}^\dagger(t_b)\mathcal{P}(t_b,t_a)\mathbf{b}_{\beta}(t_a)\boldsymbol{\Delta}_{\beta\alpha}\left(t_a, t_b\right),\nonumber\\
\end{eqnarray}
where $\alpha$ and $\beta$ are the Nambu index. Out-of-equilibrium NCA is a conserving approximation as it can be derived from the Luttinger-Ward functional~\cite{PhysRev.118.1417}. Actually, the NCA has been widely employed in the study of quenching and driven dynamics~\cite{PhysRevLett.105.146404,PhysRevB.82.115115,PhysRevB.84.035122,PhysRevX.5.011038}, as well as in the analysis of open system dynamics~\cite{PhysRevX.11.031018,10.21468/SciPostPhys.16.1.026}. We examine its application to the two-component Bose-Hubbard model in the following sections.

The self-consistency condition for the BDMFT depends on the specific lattice structure. In the subsequent discussion, we focus on the Bethe lattice, which effectively captures qualitative equilibrium characteristics of the Bose-Hubbard model in two- and three-dimensional optical lattices~\cite{PhysRevB.80.014524,Anders_2011}. We expect this validity extends to the nonequilibrium case, enabling loop closure via a simplified self-consistency relation~\cite{PhysRevX.5.011038,PhysRevX.11.031018},
 \begin{eqnarray}
 	\boldsymbol{\Delta}\left(t, t'\right) = zJ^2\mathbf{G}\left(t, t^{\prime}\right).
 \end{eqnarray}
 
The connected impurity Green’s function is given by
 \begin{equation}
 	\mathbf{G}\left(t, t^{\prime}\right)=-i\left\langle\mathcal{T_C}\mathbf{b}(t) \mathbf{b}^{\dagger}\left(t^{\prime}\right)\right\rangle + i\boldsymbol{\Phi}(t)\boldsymbol{\Phi}^{\dagger}(t'),
 \end{equation}
where the components of the connected impurity Green’s function $\mathbf{G}\left(t, t^{\prime}\right)$ are defined in appendix A.

To perform the actual calculations, we work under the Fock basis $|N_1, N_2 \rangle$ of the impurity site, and the nonequilibrium BDMFT equation can be solved self-consistently. Here, $N_1=0,\,1,\,\cdots,\,N_\mathrm{max}-1$, and  $N_2=0,\,1,\,\cdots,\,N_\mathrm{max}-1$ are the atom number for species $1$ and $2$, respectively. In our calculations, we truncate the Fock space at the maximum occupation number $N_\mathrm{max}=4$ for the Mott-insulating phase. We start from a guess for the states propagator $\mathcal{P}(t,t')$, update the self-energy $\mathcal{S}(t,t')$ from Eq.~(\ref{Self}), and then the Dyson equation Eq.~(\ref{dyson}) is solved on different branches of the contour. 
According to the symmetry of the propagator, only four components of $\mathcal{P}(t,t')$ need to be calculated, including the equilibrium component $\mathcal{P}^{M}(\tau)$, greater component $\mathcal{P}^{>}(t,t')$, right-mixing component $\mathcal{P}^{\rceil}\left(t, \tau^{\prime}\right)$, and lesser component $\mathcal{P}^{<}(t,t')$. Details can be found in Appendix A. After obtaining $\mathcal{P}(t,t')$, the connected impurity Green's function can be calculated from $\mathcal{P}(t,t')$, and then we update the hybridization function $\boldsymbol{\Delta}\left(t, t'\right)$. These procedures are repeated until convergence.

\begin{figure}[tbp]
\includegraphics[width=1.0\linewidth]{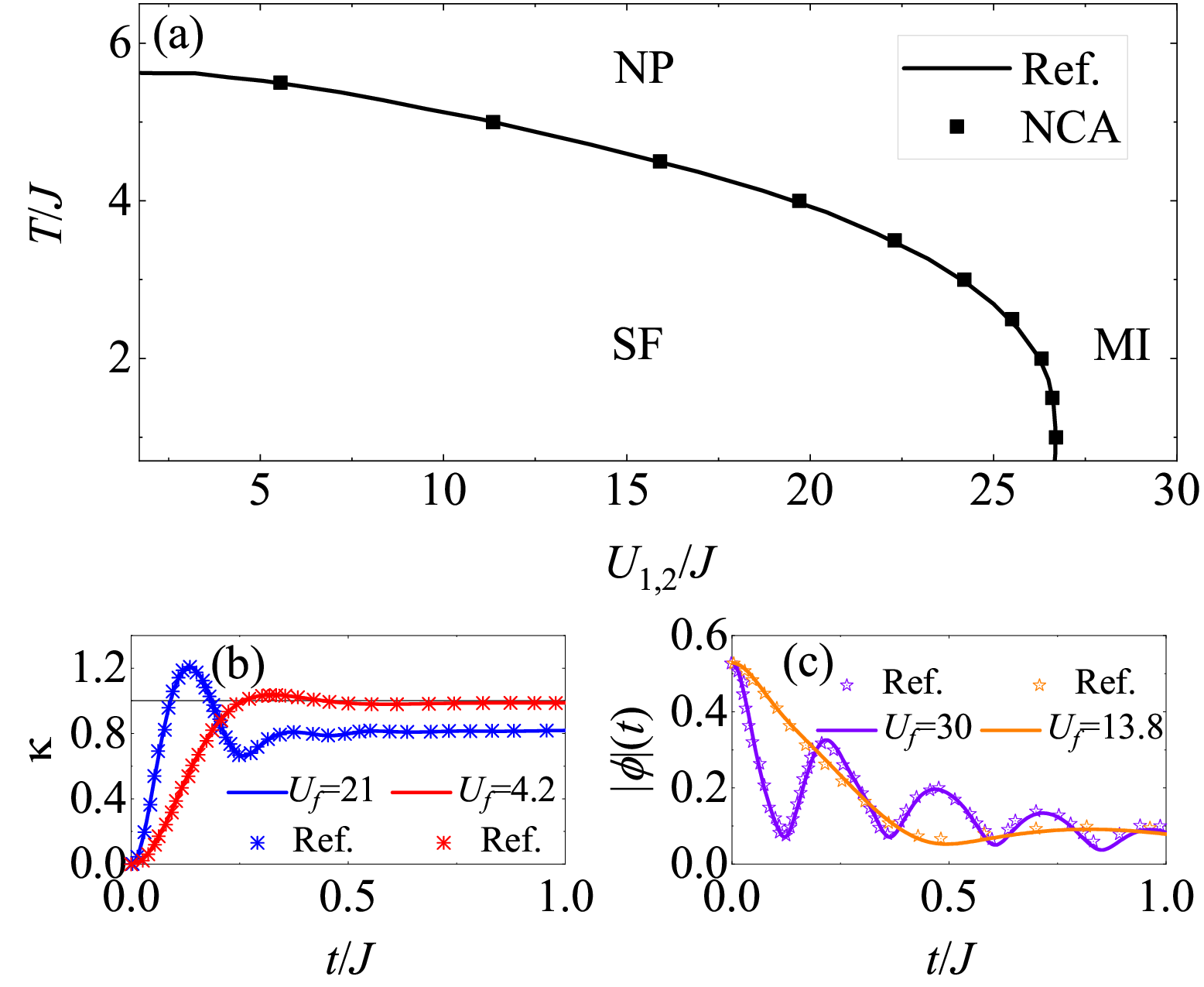}
\caption{Benchmark calculations for the decoupled two-component Bose-Hubbard model. (a) Equilibrium phase diagram on the $(U_{1,2}, T)$ plane for $U_{12}=0$, identifying superfluid (SF), normal (NP), and Mott-insulating (MI) phases. (b) Time evolution of the relative change $\kappa$ for quenching from a Mott insulator with $U_{1,2}/J=30$ and $T/J=6$ to $U_f$. (c) Time evolution of the superfluid order parameter for quenching from the superfluid phase with $U_{1,2}/J=6$ and $T/J=5.1$ to $U_f$. Our results are consistent with reported results in Ref.~\cite{PhysRevX.5.011038}. Here, we choose the filling $n_{1,2}=1$, and the Bethe lattice with infinite connectivity. 
}
\label{figure1}
\end{figure}
\section{\label{sec:level3} Results}

\subsection{Benchmark calculations for the single-component Bose-Hubbard model}
First, we verify the convergence of the impurity solver in the equilibrium properties and nonequilibrium dynamics. Actually, the NCA solver has been applied to study the single-component Bose-Hubbard model~\cite{PhysRevX.5.011038}, providing a good approximation for calculating both equilibrium and nonequilibrium phase diagrams. To validate our NCA solver implementation based on the states propagator method, we set the inter-species interaction  $U_{12}=0$, essentially decoupling Eq.~(\ref{Ham}) to two independent single-component Bose-Hubbard models. 

Both equilibrium and dynamical results are compared against those from Ref.~\cite{PhysRevX.5.011038}.
The equilibrium quantum phase diagram in the $(U_{1,2}, T)$ plane at unit filling with $\langle n_{1,2}\rangle=1$ is shown in Fig.~\ref{figure1}(a), where our results agree well with those from Ref.~\cite{PhysRevX.5.011038}. After validating against equilibrium benchmarks, we investigate out-of-equilibrium dynamics in the single-component bosonic system. 
Initializing from a deep Mott-insulating state ($U_{1,2}/J=30$ and $T/J=6$), we quench the Hamiltonian to $U_{1,2}/J\equiv U_f$ for $t>0$. 
To monitor the dynamics,  we track the relative change $\kappa$ in double occupancy $\langle n_{1,2}^2\rangle$,
\begin{eqnarray}
	\kappa = \frac{\langle n_{1,2}^2\rangle(t) -\langle n_{1,2}^2\rangle(t=0)}{\langle n_{1,2}^2\rangle_{U_f,T_\mathrm{eff}} -\langle n_{1,2}^2\rangle(t=0)},
\end{eqnarray}
for different final interactions $U_f$ [Fig.~\ref{figure1}(b)], where $T_\mathrm{eff}$ denotes the effective temperature corresponding to the initial state energy. 
In addition, we benchmark quenches from a superfluid initial state ($U_{1,2}/J=6$ and $T/J=5.1$). The evolution of the superfluid order parameter $\phi = \langle b_{1,2} \rangle$ is shown in Fig.~\ref{figure1}(c). All the calculated nonequilibrium results demonstrate excellent agreement with established literature.

\begin{figure}[tbp]
	\includegraphics[width=1\linewidth]{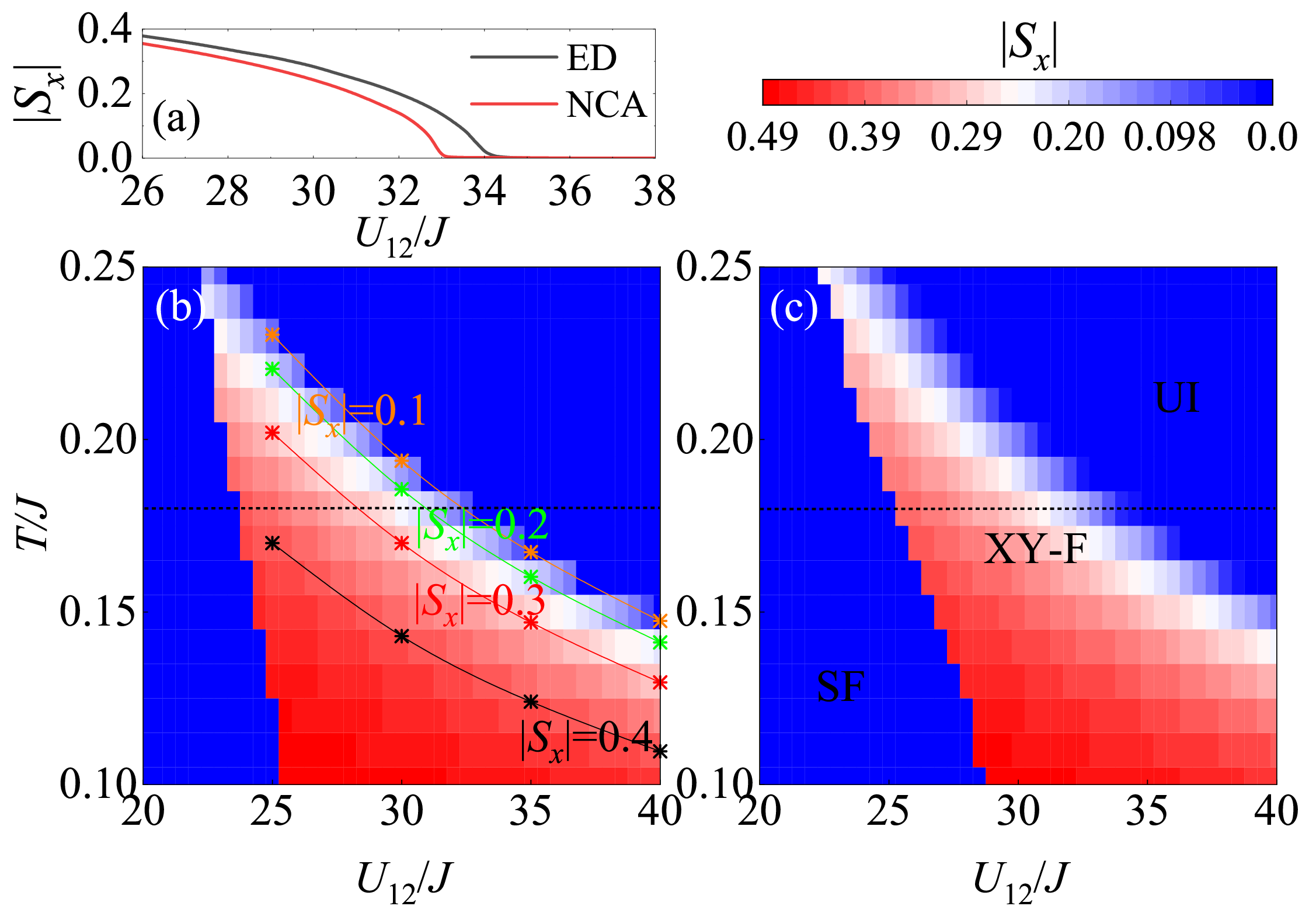}\caption{Equilibrium magnetic phase diagram of the two-component Bose-Hubbard model on the $(U_{12}, T)$ plane for the Bethe lattice with $z=6$. (a) Spin order $|S_x|$ as a function of inter-species interaction $U_{12}$, calculated by the NCA and ED solvers, respectively [black dot lines in (b) and (c)].  (b) NCA-calculated phase diagram, with contour lines are plotted for $|S_x|=0.4$, 0.3, 0.2, and 0.1, respectively. (c) ED-calculated phase diagram from Hamiltonian approaches. These two different solvers yield similar rich phases, including superfluid (SF), $xy$-ferromagnet (XY-F), and unordered insulator (UI). 
    Note here that the spin order in the SF is artificially set to zero to emphasize the XY-F and UI phases. The intra-species interaction $U_{1,2}/J=45$, and filling $n_1+n_2=1$. }
	\label{figure2}
\end{figure}

\subsection{\label{sec:level3B} Benchmark calculations for the two-component Bose-Hubbard model}
In this part, we study the phase diagram of the two-component Bose-Hubbard model, based on the NCA solver. As an extensively studied model, the equilibrium quantum phase diagrams of the two-component Bose-Hubbard model have been obtained~\cite{altman_phase_2003,PhysRevLett.91.090402,PhysRevLett.90.100401,zheng_counterflow_2025,PhysRevLett.125.245301,altman_phase_2003,PhysRevB.80.245109,PhysRevB.77.184505,PhysRevA.82.021601,PhysRevB.84.144411,PhysRevA.85.023624,li_anisotropic_2013}. Compared to the single-component one, the two-component Bose-Hubbard model supports a rich magnetic phase diagram with various spin orderings. In this work, we mainly focus on the deep Mott-insulating regime.

\begin{figure}[tbp]
	\includegraphics[width=1\linewidth]{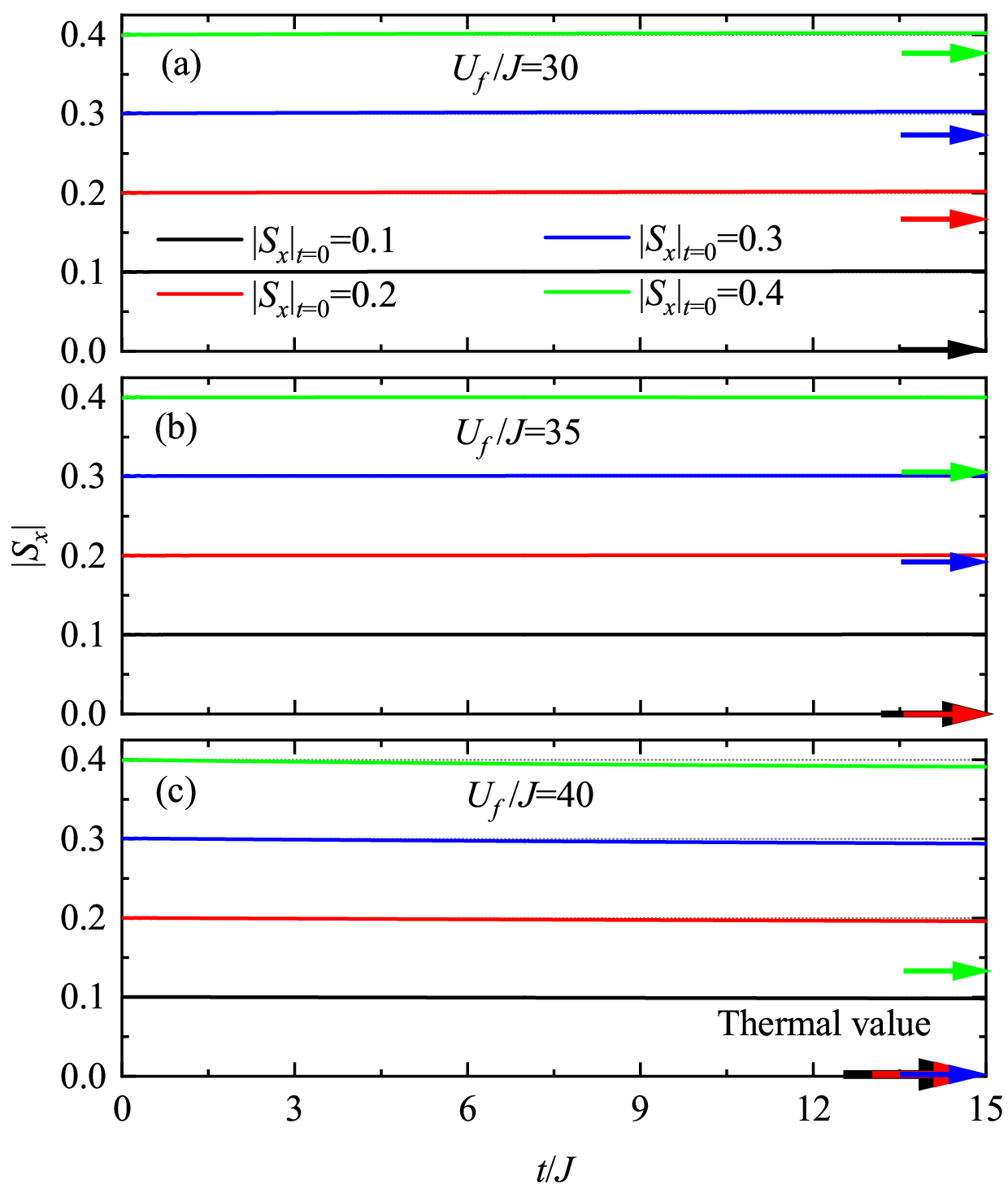}
	\caption{Real-time evolution of $|S_x|$ after quenching the inter-species interaction $U_{12}$ from $U_i/J=25$ to (a) $U_f/J=30$, (b) 35, and (c) 40 for different magnetizations. The predicted thermal values are represented by the arrows labeled with corresponding colors. Calculations are performed for the Bethe lattice with $z=6$, $U_{1,2}/J=45$, and filling $n_1+n_2=1$.}
	\label{figure3}
\end{figure}

To characterize the magnetic phases, the spin order is introduced 
${\bf \hat{S}}_i = \hat{b}^\dagger_{i,\sigma} {\bf F}_{\sigma\sigma'} \hat{b}_{i,\sigma'} $, where
${\bf F}_{\sigma\sigma^\prime}$ is the matrix for spin-1/2 particles, {\it i.e.} $\hat{S}_i^x= 1/2  ({\hat{b}_{i,1}}^\dagger {\hat{b}_{i,2}} + {\hat{b}_{i,2}}^\dagger {\hat{b}_{i,1}}) $, $\hat{S}_i^y=i/2 (-{ \hat{b}_{i,1}}^\dagger { \hat{b}_{i,2}} + {\hat{b}_{i,2}}^\dagger {\hat{b}_{i,1}}) $, and $\hat{S}_i^z= 1/2  ({\hat{b}_{i,1}}^\dagger {\hat{b}_{i,1}} - {\hat{b}_{i,2}}^\dagger  {\hat{b}_{i,2}} )$.
Based on the NCA impurity solver within BDMFT, we compute the equilibrium magnetic phase diagram in the $(U_{12}, T)$ plane through the spin order parameter $|S_x| = |\langle \hat{S}_i^x \rangle|$. We focus on the strongly interacting regime with $U_{1,2}/J = 45$, and the mass-balanced case with a fixed filling $\langle n_1\rangle + \langle n_2\rangle = 1$. Fig.~\ref{figure2}(b) presents the results from the NCA solver, while Fig.~\ref{figure2}(c) shows the corresponding phase diagram obtained via the ED solver using the Hamiltonian approach. Both methods identify identical phases in the strongly interacting regime, including superfluid (SF), $xy$-ferromagnet (XY-F), and unordered insulator (UI). Slight discrepancies in phase boundaries arise between the two numerical methods, as demonstrated in Fig.~\ref{figure2}(a), due to the NCA solver being a low-order approximation.

Our numerical results demonstrate that despite its low-order approximation, the NCA impurity solver accurately captures equilibrium magnetic properties of the two-component Bose-Hubbard model. This capability indicates that NCA should also qualitatively reproduce dynamical features of the strongly interacting bosonic system, which we will address in the following.

\subsection{\label{sec:level3C} Quench dynamics of the two-component Bose-Hubbard model}
We now examine the quench dynamics in the two-component Bose-Hubbard model on the Bethe lattice, focusing on magnetic evolution from initial equilibrium ground states. 
In the following studies, initial conditions are systematically selected along the contour lines of Fig.~\ref{figure2}(b), corresponding to fixed spin orders $|S_x|_{t=0}=0.4$, 0.3, 0.2, and 0.1, respectively. For each contour, we tune inter-species interactions 
$U_{12}$ and temperatures $T$ to satisfy these initial magnetizations. Without loss of generality, we choose the intra-species interaction $U_{1,2}/J=45$. To monitor the magnetic thermalization dynamics, we focus on the decay of the spin operator ${\hat S}_x$. Generally, the operator is expected to decay exponentially, as the non-commutation $[{\hat S}_x, {\hat H}]\neq0$ indicates that ${\hat S}_x$ is not a conserved quantity of the chaotic quantum system~\cite{PhysRevB.106.224310,mcculloch2025subexponentialdecaylocalcorrelations}.

We first study the magnetic dynamics in the strongly interacting regime by starting from an initial state with $U_{12}(t=0)\equiv U_i=25J$ to $U_{12}(t>0)\equiv U_f>U_i$.  
Fig.~\ref{figure3} shows the $|S_x|$ evolution after quenching to $U_f/J=30$, 35, and 40, respectively, where the color-matched arrows indicate the predicted thermal values of the post-quench Hamiltonian ($U_f$) at the effective temperatures associated with the initial equilibrium states. Although higher-dimensional bosonic systems are generally expected to relax toward equilibrium, we observe that the system is trapped in long-lived metastable states in the strongly interacting regime. This metastable region persists universally across all $U_f > U_i$. Crucially, in the deep Mott-insulating regime, relaxation times exhibit a substantial increase as a function of interaction strength, analogous to the behavior observed in fermionic systems where the quenches to stronger interactions lead to trapping in nonthermal states~\cite{PhysRevB.86.205101}.

\begin{figure*}[tbp]
	\includegraphics[width=1\linewidth]{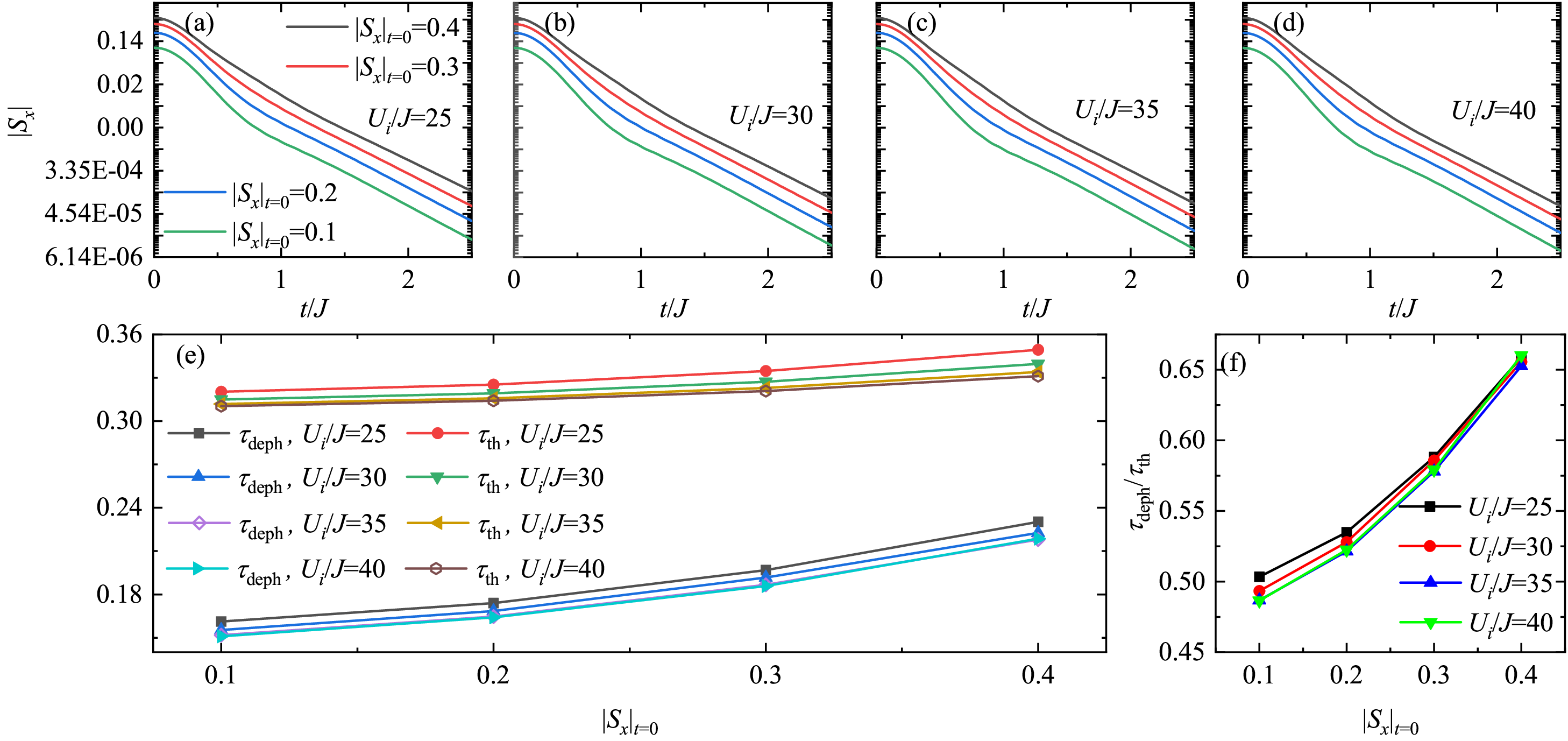}
	\caption{Two-step relaxation dynamics of $|S_x|$ following quenches to $U_{f}/J=0$ from (a) $U_i/J=25$, (b) 30, (c) 35, and (d) 40, respectively. The initial states are selected from the contour lines in Fig.~\ref{figure2}(b), with temperatures adjusted to fix the initial values $|S_x|_{t=0}=0.4$, $0.3$, $0.2$, and $0.1$ for each curve. The fitted relaxation times $\tau_\mathrm{deph}$ and $\tau_\mathrm{th}$ are plotted in (e), while their ratio is plotted in (f). Calculations are performed for the Bethe lattice with $z=6$, $U_{1,2}/J=45$, and filling $n_1+n_2=1$.}
	\label{figure4}
\end{figure*}
To relax the constraints on dynamical processes and trigger magnetic evolution, we abruptly quench the inter-species interaction from finite $U_i$ to a smaller value, such as $U_f=0$, whose Hamiltonian is associated with an unordered normal ground state~\cite{PhysRevB.80.245109}. In contrast to the previous observations (Fig.~\ref{figure3}), this protocol reduces relaxation times, enabling full thermalization of local magnetism to the equilibrium state within numerically accessible timescales.
As shown in Figs.~\ref{figure4}(a-d), we find the spin magnetization $|S_x|$ decreasing exponentially towards the thermal equilibrium values. Interestingly, we find the dynamics of $|S_x|$ exhibiting a two-step relaxation with distinct short- and long-timescale decay rates. We extract the relaxation times  $\tau_\mathrm{deph}$ (short timescale) and $\tau_\mathrm{th}$ (long timescale) by fitting the evolution of $|S_x|$ to $e^{-t/\tau}$ separately for each regime.

The obtained $\tau_\mathrm{deph}$ and $\tau_\mathrm{th}$ are plotted against initial magnetization $|S_x|_{t=0}$ in Fig.~\ref{figure4}(e). We find  $\tau_{\mathrm{th}}$ remains nearly constant across all initial magnetizations $|S_x|_{t=0}$, while $\tau_\mathrm{deph}$ grows with increasing $|S_x|_{t=0}$ for all the discussed interaction strengths. This behavior is consistent with thermalization processes in the chaotic quantum systems: after a rapid initial relaxation, which is sensitive to the initial state, the system enters into a slower thermalization stage independent of the initial state. 
To see it more clearly, the ratio $\tau_\mathrm{deph}/\tau_\mathrm{th}$ is shown in Fig.~\ref{figure4}(f), where it becomes smaller with decreasing the initial magnetization $|S_x|_{t=0}$, indicating a more pronounced two-step relaxation.

\subsection{\label{sec:level3D} Magnetic dynamics induced by periodic modulation of $U_{12}$}
In the following, we investigate the Floquet dynamics of Eq.~(\ref{Ham}), 
focusing on the time-dependent variation of inter-species interaction $U_{12}$,
\begin{eqnarray}
	U_{12}(t) = U_i\left[1+\alpha \sin(\omega t) \right],
\end{eqnarray}
where $\alpha$ is the relative drive strength, and $\omega$ is the drive frequency. Here, we start from an equilibrium state for $U_i/J=25$ and $U_{1,2}/J=45$.
\begin{figure}[tbp]
	\includegraphics[width=1.0\linewidth]{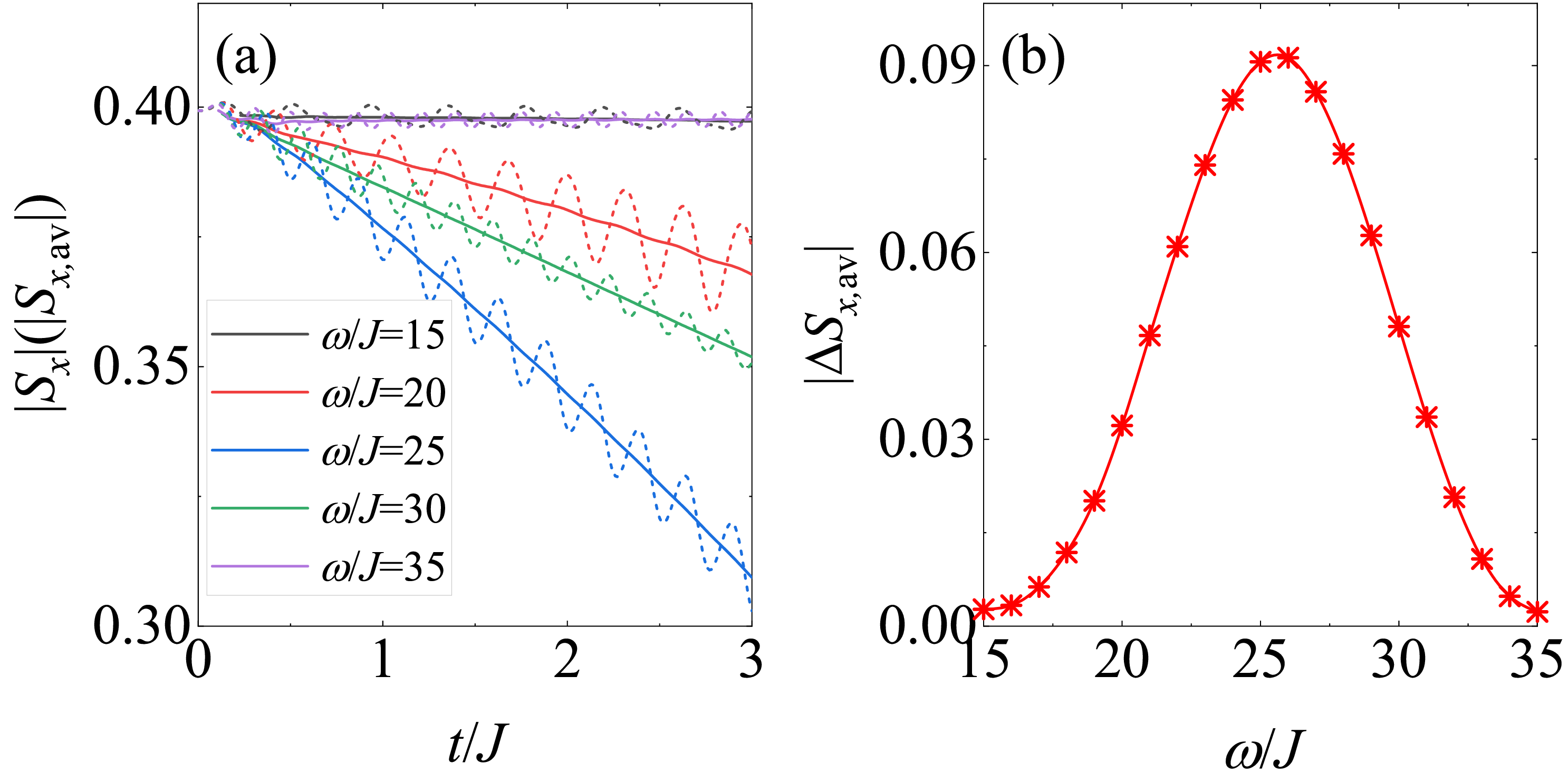}
	\caption{(a) Short-time evolution of $|S_x(t)|$ and $|S_{x,\mathrm{av}}(t)|$ for a periodic modulation of $U_{12}$ at different drive frequencies $\omega$. The dot lines denote the real-time evolution $|S_x(t)|$, and the solid ones its moving average $|S_{x,\mathrm{av}}(t)|$. (b) Relative change of $|S_{x,\mathrm{av}}(t)|$ as a function of $\omega$. Calculations are performed for the Bethe lattice with $z=6$, other parameters are $\alpha=0.2$, $U_{1,2}/J=45$, $U_i/J=25$, and filling $n_1+n_2=1$.}
	\label{figure5}
\end{figure}

First, we study the short-time magnetic dynamics driven by modulating the inter-species interactions. In Fig.~\ref{figure5}(a), we demonstrate the real-time evolution $|S_x(t)|$ (dot line) and its moving average $|S_{x,\mathrm{av}}(t)|=\frac{\omega}{2\pi}\int_{t-\pi/\omega}^{t+\pi/\omega}\mathrm{d}t^\prime|S_x(t^\prime)|$ (solid line) during a periodic modulation of $U_{12}$ for diverse driving frequencies $\omega$. As expected, the moving average closely tracks the instantaneous Floquet dynamics.
In addition, we find that the averaged spin order $|S_{x,\mathrm{av}}|$ exhibits rapid decay when $\omega \approx U_i$ but slower evolution for the off-resonance drives. To verify this point, relative change of $|\Delta S_{x,\mathrm{av}}|=|S_{x,\mathrm{av}}(t_i)-S_{x,\mathrm{av}}(t_f)|$ for $t_i/J=0$ and $t_f/J=3$ is shown in Fig.~\ref{figure5}(b), revealing a Floquet spectrum with a peak at $\omega \approx U_i$.

We next examine the long-time modulation dynamics of magnetic order. To resolve the rapid decay of finite  $|S_x|$ to zero, we fix the driving frequency at $\omega = U_i$ [Fig.~\ref{figure5}(b)]. Under sustained Floquet driving, persistent energy injection heats the many-body system. This heating eventually leads to thermal fluctuations that destroy the long-range spin order, driving a crossover from the XY-F phase to the UI phase.

\begin{figure}[tbp]
	\includegraphics[width=1.0\linewidth]{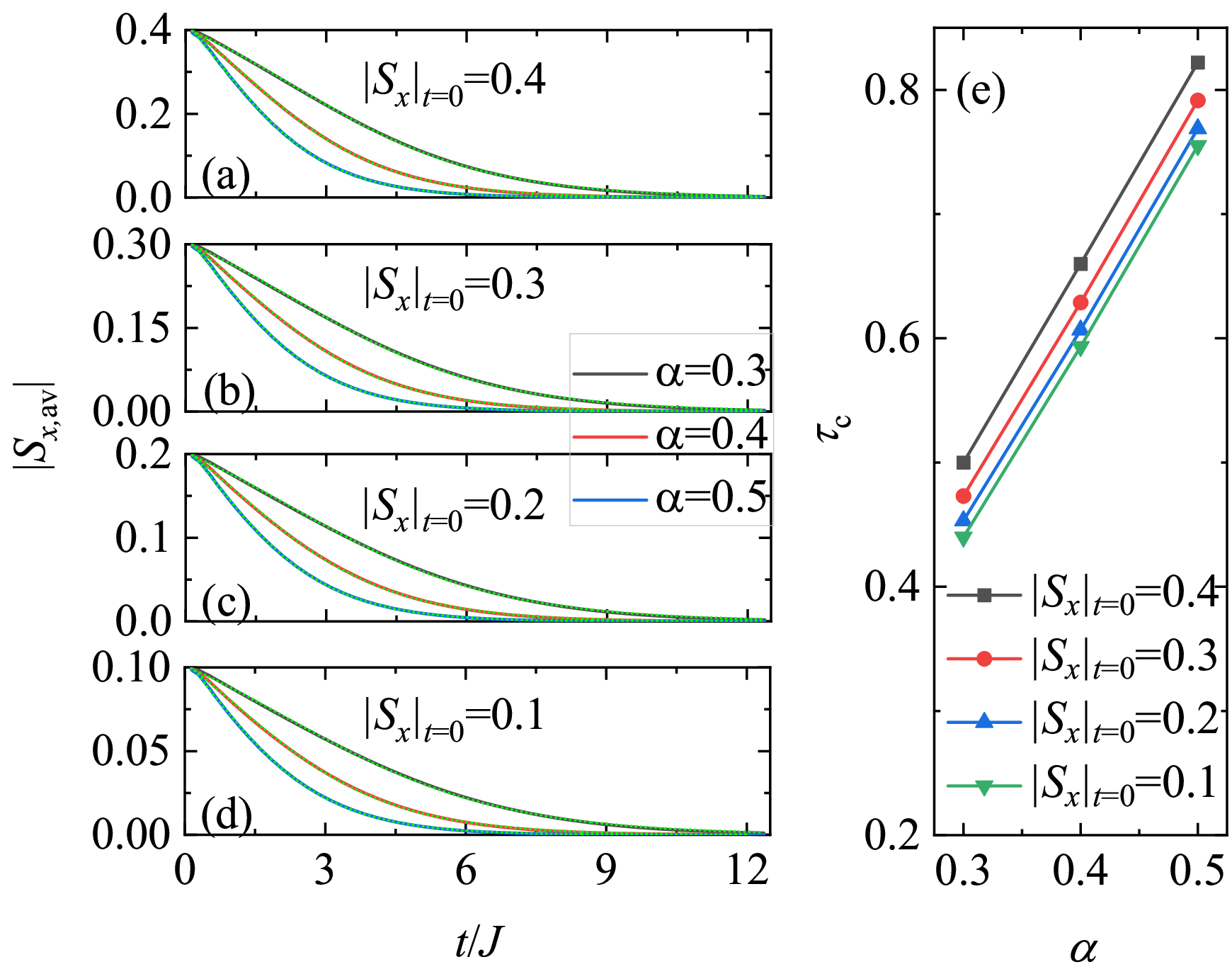}
	\caption{Long-time evolution of $|S_{x,\mathrm{av}}(t)|$ for a periodic modulation of $U_{12}$ at different initial transverse magnetizations (a) $|S_x|_{t=0} = 0.4$, (b) 0.3, (c) 0.2, and (d) $0.1$  (solid lines). The dashed lines are a sigmoid fit of $|S_{x,\mathrm{av}}(t)|$ for three different drive strengths $\alpha$. (e) The fitted $\tau_c$ as a function of $\alpha$. Calculations are performed for the Bethe lattice with $z=6$, $U_{1,2}/J=45$, $U_i/J=25$, and filling $n_1+n_2=1$.}
	\label{figure6}
\end{figure}
Figs.~\ref{figure6}(a-d) show the evolution of the averaged $|S_{x,\mathrm{av}}|$ for different modulation strengths $\alpha$ and various initial transverse magnetizations $|S_x|_{t=0}$.
Interestingly, we find that the evolution of the averaged $|S_{x,\mathrm{av}}|$ under modulation conforms to a sigmoid function
\begin{eqnarray}
	|S_{x,\mathrm{av}}(t)|=S_0\left[1-\frac{1}{1+e^{-\tau_c(t-t_0)}} \right],
\end{eqnarray}
as fitted by the green dashed lines in Figs.~\ref{figure6}(a-d). Here, $S_0$, $t_0$, and $\tau_c$ denote the fitting parameters. In addition, Fig.~\ref{figure6}(e) illustrates the variation of the factor $\tau_c$ with respect to $\alpha$, which increases linearly as a function of $\alpha$.

\section{\label{sec:level4} Conclusions}
We extend the established nonequilibrium bosonic dynamical mean-field theory from the single- to two-component systems, utilizing the noncrossing approximation impurity solver. 
Validation against the single-component Bose-Hubbard model shows excellent agreement with previous results for both ground states and quench dynamics.
For the two-component bosonic case, the ground-state magnetic phase diagrams align with the results from the Hamiltonian-based exact diagonalization solver. 
Crucially, quench dynamics reveals that rapidly increasing inter-species interactions induces slow thermalization toward long-lived metastable states.
Conversely, reducing inter-species interactions to zero results in rapid thermal equilibrium via a unique two-step exponential relaxation. Additionally, the study investigates Floquet dynamics arising from periodic modulation of inter-species interactions, unveiling magnetic-to-disordered phase transitions. For future work, it will be interesting to investigate nonequilibrium phase diagrams following quenches from the superfluid phase, nonequilibrium phase transitions for the mass-imbalanced systems, and the dynamics associated with attractive inter-species interactions.

\section{\label{sec:level7} Acknowledgements} 
We acknowledge useful discussions
with Tao Qin and H. U. R. Strand. This work is supported by the National Natural Science Foundation of China under ation of China under Grants Nos. 12374252, 12074431, and 12174130, and
the Science and Technology Innovation Program of Hunan
Province under Grant No. 2024RC1046. We acknowledge the ChinaHPC for providing HPC resources that have contributed to the research results reported within this paper.

\begin{widetext}
	\begin{center}
		 \section{\label{sec:level8} Appendix}
	\end{center}
	\renewcommand{\theequation}{A\arabic{equation}}
	\renewcommand{\thesection}{S-\arabic{section}}
	\renewcommand{\bibnumfmt}[1]{[S#1]}
	\renewcommand{\citenumfont}[1]{S#1}
	\setcounter{equation}{0}

\subsection{\label{sec:level8A} Numerical solution of Volterra integro-differential equation for states propagator}

We start from the Dyson equation for states propagator
\begin{eqnarray}
	i \partial_t \mathcal{P}\left(t, t^{\prime}\right)=\mathcal{H}_{\mathrm{loc}}(t) \mathcal{P}\left(t, t^{\prime}\right)+\int_{t^{\prime}}^t \mathrm{d} t_a \mathcal{S}\left(t, t_a\right) \mathcal{P}\left(t_a, t^{\prime}\right).
\end{eqnarray}
Here, the convolution of self-energy and propagator $\int_{t^{\prime}}^t \mathrm{d} t_a \mathcal{S}\left(t, t_a\right) \mathcal{P}\left(t_a, t^{\prime}\right)=\int_{t}^{t'}\mathrm{d}t_ag(t_a)$ is defined as~\cite{RevModPhys.86.779}
\begin{eqnarray}
	\int_{t}^{t'}\mathrm{d}t_ag(t_a) = \left\{ \begin{array}{lcl}
			\int_{t}^{t'}\mathrm{d}t_ag(t_a) & \mbox{if\ } t\succ t',\\
			 \int_{0}^{t}\mathrm{d}t_ag(t_a) + \int_{t'}^{-i\beta}\mathrm{d}t_ag(t_a) & \mbox{if\ } t\prec t',
		\end{array}\right.
\end{eqnarray}
which can be expressed by using the Langreth rules~\cite{langreth_linear_1976,Stefanucci_van_Leeuwen_2013}. Then, we derive the following four coupled integral equations
\begin{eqnarray}
	-\partial_\tau \mathcal{P}^M(\tau)=\mathcal{H}_{\mathrm{loc}} \mathcal{P}^M(\tau)+\int_0^\tau \mathrm{d} \bar{\tau} \mathcal{S}^M(\tau-\bar{\tau}) \mathcal{P}^M(\bar{\tau}),
\end{eqnarray}

\begin{eqnarray}
	i \partial_t \mathcal{P}^{>}\left(t, t^{\prime}\right)=\mathcal{H}_{\text {loc }}(t) \mathcal{P}^{>}\left(t, t^{\prime}\right)+\int_{t^{\prime}}^t \mathrm{~d} \bar{t} \mathcal{S}^{>}(t, \bar{t}) \mathcal{P}^{>}\left(\bar{t}, t^{\prime}\right),
\end{eqnarray}

\begin{eqnarray}
	i \partial_t \mathcal{P}^{\rceil}\left(t, \tau\right)=\mathcal{H}_{\mathrm{loc}}(t) \mathcal{P}^{\rceil}\left(t, \tau\right)+\int_{0}^t \mathrm{~d} \bar{t} \mathcal{S}^{>}\left(t, \bar{t}\right) \mathcal{P}^{\rceil}\left(\bar{t}, \tau\right) + \int_{\tau}^{\beta} \mathrm{~d} \bar{\tau} \mathcal{S}^{\rceil}\left(t, \bar{\tau}\right) \mathcal{P}^{M}\left(\bar{\tau}-\tau\right),
\end{eqnarray}

\begin{eqnarray}
	i \partial_t \mathcal{P}^{<}\left(t, t^{\prime}\right)&=&\mathcal{H}_{\mathrm{loc}}(t) \mathcal{P}^{<}\left(t, t^{\prime}\right)+\int_{0}^t \mathrm{~d} \bar{t} \mathcal{S}^{>}\left(t, \bar{t}\right) \mathcal{P}^{<}\left(\bar{t}, t^{\prime}\right)\nonumber\\ &-&\int_{0}^{t^{\prime}} \mathrm{~d} \bar{t} \mathcal{S}^{<}\left(t, \bar{t}\right) \mathcal{P}^{>}\left(\bar{t}, t^{\prime}\right)-i\int_{0}^{\beta} \mathrm{~d} \bar{\tau} \mathcal{S}^{\rceil}\left(t, \bar{\tau}\right) \mathcal{P}^{\lceil}\left(\bar{\tau}, t^{\prime}\right),
\end{eqnarray}
for equilibrium, greater, right-mixing, and lesser components, respectively. Notice here, $\mathcal{P}^{M,>,\rceil,<}(t,t')$ is a $N_\mathrm{max}^2\times N_\mathrm{max}^2$ matrix under the Fock basis $|N_1, N_2 \rangle$ of the impurity site. Discretizing the imaginary-times interval $[0,\beta]$ and real-time interval $[0,t_\mathrm{max}]$, these equations can be solved numerically using a fifth-order multistep method~\cite{PhysRevX.5.011038,schuler_nessi_2020} or an implicit second-order Runge-Kutta scheme~\cite{RevModPhys.86.779,PhysRevX.11.031018,10.21468/SciPostPhys.16.1.026}, we adopt the latter approach, and its results align well with those of the former method.
Then the connected impurity Green’s functions are calculated by
\begin{equation}
	\begin{aligned}
		\mathbf{G}^M(\tau) & =-\frac{1}{Z} \operatorname{Tr}\left[\mathcal{P}(\beta-\tau) \mathbf{b}(\tau) \mathcal{P}(\tau) \mathbf{b}^{\dagger}(\tau)\right] + \boldsymbol{\Phi}(\tau)\boldsymbol{\Phi}^{\dagger}(\tau), \\
		\mathbf{G}^{>}\left(t, t^{\prime}\right) & =\frac{i}{Z} \operatorname{Tr}\left[\mathcal{P}^{<}\left(t^{\prime}, t\right) \mathbf{b}(t) \mathcal{P}^{>}\left(t, t^{\prime}\right) \mathbf{b}^{\dagger}(t')\right] + i\boldsymbol{\Phi}(t)\boldsymbol{\Phi}^{\dagger}(t'), \\
		\mathbf{G}^{\rceil}(t, \tau) & =\frac{i}{Z} \operatorname{Tr}\left[\mathcal{P}^{\lceil}(\tau, t) \mathbf{b}(t) \mathcal{P}^{\rceil}(t, \tau) \mathbf{b}^{\dagger}(\tau)\right] + i\boldsymbol{\Phi}(t)\boldsymbol{\Phi}^{\dagger}(\tau), \\
		\mathbf{G}^{<}\left(t, t^{\prime}\right) & =\frac{i}{Z} \operatorname{Tr}\left[\mathcal{P}^{>}\left(t^{\prime}, t\right) \mathbf{b}(t) \mathcal{P}^{<}\left(t, t^{\prime}\right) \mathbf{b}^{\dagger}(t')\right] + i\boldsymbol{\Phi}(t)\boldsymbol{\Phi}^{\dagger}(t').
	\end{aligned}
\end{equation}

\subsection{\label{sec:level8B} Calculation of Energy}
The total energy of the system is expressed as follows:
\begin{eqnarray}
	E_\mathrm{tot} = E_\mathrm{kin} + E_\mathrm{con} + E_\mathrm{int},
\end{eqnarray}
where the kinetic energy reads
\begin{eqnarray}
	E_\mathrm{kin} = \frac{i}{2}\mathrm{Tr}\left[(\mathbf{\Delta}*\mathbf{G})^<(t,t)\right],
\end{eqnarray}
the condensate energy is given by
\begin{eqnarray}
	E_\mathrm{con} = -zJ\sum_{\sigma}|\langle b_{\sigma}(t) \rangle|^2,
\end{eqnarray}
and the local interaction energy is described as
\begin{eqnarray}
	E_\mathrm{int} = U_{\sigma\sigma'}\langle n_{\sigma}(t)n_{\sigma'}(t)\rangle+\sum_{\sigma}\frac{U_{\sigma}}{2}\left[ \langle n_{\sigma}(t)n_{\sigma}(t)\rangle-\langle n_{\sigma}(t)\rangle\right].
\end{eqnarray}

\end{widetext}

\bibliography{references}

\end{document}